\documentclass[sigconf]{acmart}
\usepackage{multirow}
\usepackage{enumitem}

\newcommand{\tool}{\texttt{dVermin}}

\usepackage[linesnumbered,ruled,vlined]{algorithm2e}
\AtBeginDocument{%
  \providecommand\BibTeX{{%
    \normalfont B\kern-0.5em{\scshape i\kern-0.25em b}\kern-0.8em\TeX}}}

\copyrightyear{2022} 
\acmYear{2022} 
\setcopyright{acmcopyright}\acmConference[ASE '22]{37th IEEE/ACM International Conference on Automated Software Engineering}{October 10--14, 2022}{Rochester, MI, USA}
\acmBooktitle{37th IEEE/ACM International Conference on Automated Software Engineering (ASE '22), October 10--14, 2022, Rochester, MI, USA}
\acmPrice{15.00}
\acmDOI{10.1145/3551349.3556935}
\acmISBN{978-1-4503-9475-8/22/10}

\begin{document}

\title{The Metamorphosis: Automatic Detection of Scaling Issues for Mobile Apps}

\author{Yuhui Su}
\authornote{Also With Laboratory for Internet Software Technologies, Institute of Software, CAS}
\authornote{Also With University of Chinese Academy of Sciences}
\email{su.yu.hui@icloud.com}
\affiliation{%
\institution{Institute of Software, Chinese Academy of Sciences}
\city{Beijing}
\country{China}
}

\author{Chunyang Chen}
\email{chunyang.chen@monash.edu}
\affiliation{%
  \institution{Monash University}
  \city{Melbourne}
  \country{Australia}
}

\author{Junjie Wang}
\authornotemark[1]
\authornotemark[2]
\authornote{Corresponding author}
\authornote{Also With State Key Laboratory of Computer Sciences, Institute of Software, CAS}
\email{junjie@iscas.ac.cn}
\affiliation{
\institution{Institute of Software, Chinese Academy of Sciences}
\city{Beijing}
\country{China}
}

\author{Zhe Liu}
\authornotemark[1]
\authornotemark[2]
\email{liuzhe181@mails.ucas.edu.cn}
\affiliation{%
\institution{Institute of Software, Chinese Academy of Sciences}
\city{Beijing}
\country{China}
}

\author{Dandan Wang}
\authornotemark[1]
\authornotemark[2]
\authornotemark[4]
\email{dandan@iscas.ac.cn}
\affiliation{%
\institution{Institute of Software, Chinese Academy of Sciences}
\city{Beijing}
\country{China}
}

\author{Shoubin Li}
\authornotemark[1]
\authornotemark[2]
\authornotemark[4]
\email{shoubin@iscas.ac.cn}
\affiliation{
\institution{Institute of Software, Chinese Academy of Sciences}
\city{Beijing}
\country{China}
}

\author{Qing Wang}
\authornotemark[1]
\authornotemark[2]
\authornotemark[3]
\authornotemark[4]
\email{wq@iscas.ac.cn}
\affiliation{
\institution{Institute of Software, Chinese Academy of Sciences}
\city{Beijing}
\country{China}
}


\begin{abstract}
As the bridge between users and software, Graphical User Interface (GUI) is critical to the app  accessibility.
Scaling up the font or display size of GUI can help improve the visual impact, readability, and usability of an app, and is frequently used by the elderly and people with vision impairment. 
Yet this can easily lead to scaling issues such as text truncation, component overlap, which negatively influence the acquirement of the right information and the fluent usage of the app.
Previous techniques for UI display issue detection and cross-platform inconsistency detection cannot work well for these scaling issues.
In this paper, we propose an automated method, {\tool}, for scaling issue detection, through detecting the inconsistency of a view under the default and a larger display scale. The evaluation result shows that {\tool} achieves 97\% precision and 97\% recall in issue page detection, and 84\% precision and 91\% recall for issue view detection, outperforming two state-of-the-art baselines by a large margin. We also evaluate {\tool} with popular Android apps on F-droid, and successfully uncover 21 previously-undetected scaling issues with 20 of them being confirmed/fixed.
\end{abstract}

\ccsdesc[500]{Software and its engineering}
\ccsdesc[500]{Software and its engineering~Software testing and debugging}

\keywords{Software Testing, Android Testing, UI Inconsistency Issue, Accessibility Testing}

\maketitle

\section{Introduction}

The mobile application plays a significant part in our everyday lives, providing the convenience associated with consumer use of smartphones for communications, entertainment and finance reasons.
A good GUI (Graphical User Interface)\footnote{This work uses ``GUI'' and ``UI'' interchangeably.} design makes an application easy, practical and efficient to use, which significantly affects the success of the application and the loyalty of its users. 
For example, computer users view Apple's Macintosh system as having better GUIs than Windows system; 
their positive views almost double that of Windows users, leading to 20\% more brand loyalty~\cite{10.1145/203241.203259}.

Due to the importance of mobile apps, ensuring their accessibility to a broad range of users is crucial, i.e., the app should be usable by as many people as possible, regardless of any disabilities they might have. 
As the bridge between users and software, GUI is critical to app accessibility.
The most common disabilities that affect mobile applications are vision and motor control impairments.
When users with such issues are using mobile apps, they tend to increase the font size or display size of the screen for easier manipulation.
One recent survey studying 1.5 million mobile phone users~\cite{accessibility_settings_do_the_Dutch_prefer}
found that 33\% of people change the text size on their phone (20\% increasing the font size).

\begin{figure}[h]
	\centering
	\includegraphics[width=0.65\linewidth]{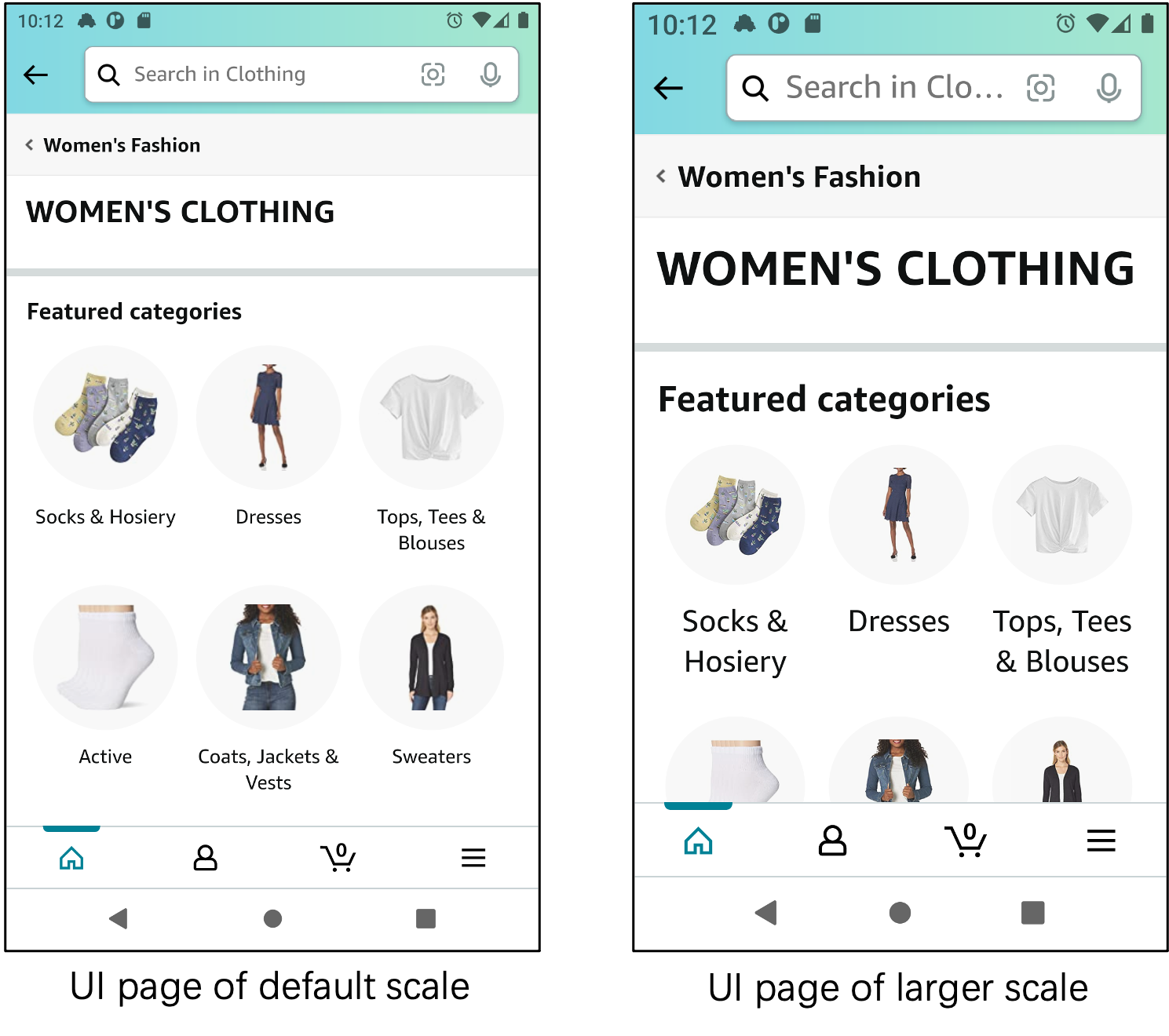}
	\vspace{-0.1in}
	\caption{Pages under the default scale setting and a larger scale setting from the ``Amazon Shpping'' app \cite{amazon}. 
	}
	\label{fig:intro}
	\vspace{-0.1in}
\end{figure}

It especially applies to the elderly, as everyone starts to feel the effects of presbyopia from middle age.
However, scaling up the font or display size of GUI can help improve visual impact (e.g., messages become louder, confident), readability (even from a distance), and usability (e.g., minimise clutter, reduce extraneous cognitive load) as seen in Figure~\ref{fig:intro}.
With a larger font size, the elderly ($>$65), despite with declining vision, hearing, and cognition, can still perform as well as young people in using mobile apps effectively~\cite{YEH2020e04147}.

In addition to users with vision impairment (e.g., low vision, far/nearsightedness), general users with eyestrain or under strong sunshine may also scale up the font or display size for a clear view of mobile GUI.

However, the GUI of many apps is only designed for  the default setting of the font or display size, as
most developers are always with good eyesight, who were not in the shoes of users with vision impairment.
This leads to the GUI inconsistency when scaling up from the default setting to larger font or display size.
In particular, the UI may be distorted with issues such as text truncation (e.g., half a line or the end of the text is missing), overflow components (e.g., the component has grown beyond the boundary of the screen), component overlap (e.g., an icon overlaying the text as the text becomes longer due to resizing), crappy layout (component at the top was placed to the bottom), etc.
A more detailed analysis of common scaling issues can be seen in our empirical study (Section~\ref{UIAutormator_and_Layout_Inspector}).
Even if responsive frameworks like reactjs \cite{reactjs} are used, it may still not work well when the size changes dramatically.
Although the app can still run along with these issues, they negatively influence the acquirement of the right information and the fluent usage of the app, resulting in a significantly bad user experience and corresponding loss of users.

Although there are already some official guidelines about ensuring text size and weight of the GUI by Google and Apple \cite{apple_design}, it is too abstract without tool support in the current practice of automatic UI testing for detecting the scaling issues. 
Manually testing these issues involves manually exploring the app, recording the screenshot under the default display scale, and the counterpart screenshot under a larger display scale, comparing each component and checking the inconsistencies, which is time- and effort-consuming. 
Existing approaches for automatically detecting UI display issues, e.g., OwlEyes \cite{DBLP:conf/kbse/LiuCWHHW20}, which takes as input the screenshot of an app, applies visual understanding technique for spotting noticeable UI display issues, is not able to detect the scaling issues about content missing.
DiffDroid \cite{8115644} detects the cross-platform inconsistency by extracting the UI component pairs with Android component id and XPath, and conducting the inconsistency detection per component pair. 
Yet the low precision of both component pair extraction and structural/color similarity detection of images limit its application in our scaling issue detection.

Nevertheless, motivated by the cross-platform inconsistency detection, we assume the scaling issues can also be detected by comparing the appearance of a UI page under the default display scale and under a larger display scale, and detecting the inconsistency among them. 
Yet there are the following challenges for this task. 
First, directly comparing two UI pages for scaling issue detection can easily bring much noise, and how to accurately build the corresponding relationship between two UI elements under different scales is challenging. 
Second, the components within a UI page can interact with each other, and when using a larger display scale, the interaction effect makes the issue detection more complex.  
Third, when using a larger display scale, different types of content (e.g., text, image) in a view might be scaled differently; thus the inconsistency detection should consider the characteristics of different types of views. 

To overcome these challenges, we propose an automated approach for the scaling issue detection, which detects the inconsistency of a view under default display scale and larger display scale. 
The scaling issue is triggered by changing the display scale of an app, which is quite similar to the tragedy which happened to the protagonist salesman, Gregor Samsa, in ``\textbf{The Metamorphosis}'' \cite{kafka1948metamorphosis}, a Franz Kafka's famous fiction, who inexplicably changed into a huge vermin and eventually died grievously.
Considering this, we name our approach as {\tool} (for detecting vermin) more vividly. 

There are three phases in {\tool}.
The first phase is view pairing. 
For a UI page under the default display scale and its counterpart UI page under a larger display scale, the first phase maps each view with its counterpart view for better conducting the follow-up scaling detection. 
The second and third phase checks the scaling issue from the perspective of the interaction between views and single view, respectively. 
In detail, the second phase is inter-view inconsistency detection, which detects whether there is any overlapping/cropping between two views in a UI page, and checks whether the overlapping/cropping status remains consistent  for two pages.
The third phase is intra-view inconsistency detection, which checks the inconsistency for each view and its counterpart view.

To evaluate the effectiveness of our {\tool}, we experiment with 60 apps involving 96 buggy UI pages and 147 buggy views. {\tool} can achieve 97\% precision and 97\% recall for issue page detection, and achieve 84\% precision and 91\% recall for issue view localization, outperforming the two state-of-the-art baselines by a large margin. 
Apart from the accuracy of our {\tool}, we also evaluate the usefulness of {\tool} by applying it in detecting the scaling issues in real-world apps from F-Droid. 
We find 21 scaling issues and report them to the development team, among which 20 (95\%) of them are confirmed/fixed by the developers, with one pending. 

The contributions of this paper are as follows:
\begin{itemize}[leftmargin=10pt]
    \item The first work focusing on the detection of scaling issues of mobile apps, to the best of our knowledge. We begin with a pilot study to reveal the prevalence of the scaling issues induced by using a larger display scale and categorize the issues for facilitating follow-up studies.
    \item The first automated approach {\tool} for mobile app scaling issue detection, which contains three phases including view pairing, inter-view inconsistency detection, and intra-view inconsistency detection to check the inconsistency of a view under default and larger display scales.
    \item Effectiveness evaluation of {\tool} with 60 apps achieving high precision and recall of scaling issue detection; and usefulness evaluation of {\tool} with 20 confirmed/fixed GitHub issue reports\footnote{We release the source code, dataset, and detailed experimental results in our website \hyperlink{https://github.com/dVermin/dVermin}{https://github.com/dVermin/dVermin}  to facilitate the replication and follow-up studies.}.
\end{itemize}

\section{Background}\label{background}

\subsection{View and View Tree (VT) in Android}\label{view_in android}

All elements in a UI page are built using a hierarchy of \textit{View} and \textit{ViewGroup} objects  \cite{ui_Layout}.
A \textit{View} usually draws something the user can see and interact with, whereas a \textit{ViewGroup} is an invisible container that defines the layout structure for \textit{View} and other \textit{ViewGroup} objects  \cite{ui_Layout}.  
The \textit{View} objects are usually called ``widgets'' and can be one of many subclasses, such as \textit{Button} or \textit{TextView}.
The \textit{ViewGroup} objects are usually called ``layouts'' and can be one of many types that provide a different layout structure, such as \textit{LinearLayout} or \textit{ConstraintLayout}.
The structure of all views in a UI page is called the layout \cite{ui_Layout}. 
Since they are organized into a single tree, we call it \textit{View Tree (short for VT)}.

\begin{figure}[t]
  \centering
  \includegraphics[width=0.75\linewidth]{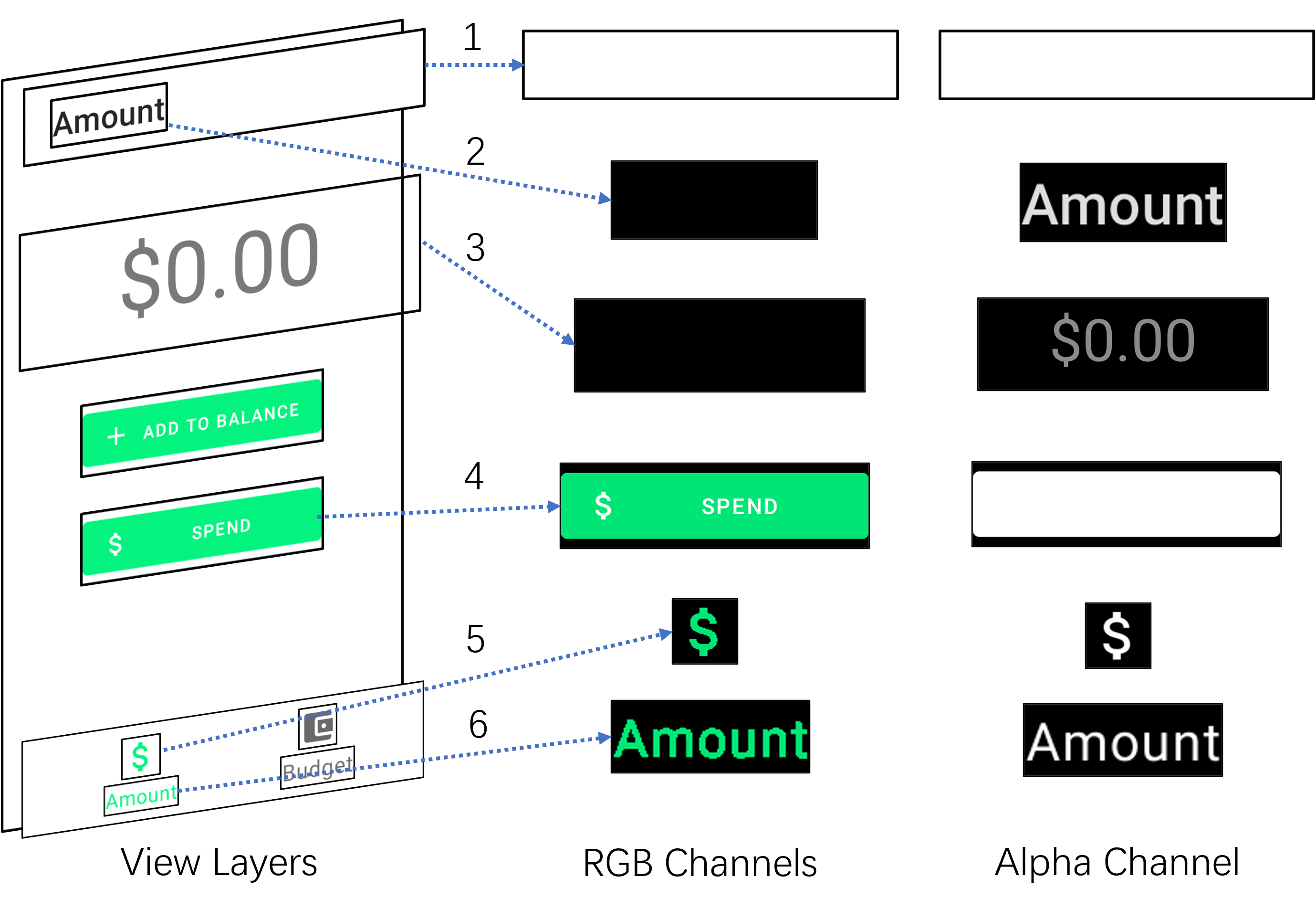}
  \caption{Layers of view and RGB channels, the alpha channel of images for example views. Each view is surrounded with a black border for clarification.}
  \label{fig:layers}
  \vspace{-0.1in}
\end{figure}

During the GUI rendering, the screen can be seen as a canvas to draw on. 
Each view can be treated as an image containing two parts of the same size, i.e., image of RGB channels and image of alpha channel in Figure \ref{fig:layers}.
Values of an alpha channel are in the range between 0 and 255, telling Android the percentage of the view should be kept when blending it with others through skia \cite{skia}. 
Each view's image contains images of itself and all of its off-spring views. 
Inside a view tree, the parent view is drawn before its children, and the children are drawn above their parent according to their z-order values.
The child with the lower z-order will be drawn first. For those with the same z-order, they will be drawn according to their pre-order traversing sequence\cite{drawing}.

When drawing a view, Android treats the image of its parent view as the destination $Dest$,
and image of itself as the source $Src$. Based on the alpha channel of $Src$, the resulted image, $Res$, is drawn according to
$Res = Src + (255-Src_\alpha)/255\times Dest$ by default \cite{kSrcOver,canvas_in_android}. $Src_\alpha$ is the alpha channel of $Src$, and each channel of $Dest$ and $Src$ are blended in this way, forming each channel of $Res$.

Scale related settings, including the display size (also known as zooming size) and font size, can manipulate the size of a view displayed on the screen. Display size is for scaling views specified with density-independent pixels or scale-independent pixels. 
Font size is for scaling text related views (e.g., TextView) specified with scale-independent pixels. 
When either of the two settings is changed, the size of the views could change \cite{density_independent,dimension}.

\subsection{Collecting View Tree (VT)}
\label{UIAutormator_and_Layout_Inspector}

For debugging, one commonly used tool is UIAutomator for generating an UI hierarchy for an app, including bounding box, accessibility properties of UI components and the layout structure \cite{UIAutomator}. 
It is useful in such tasks as automated testing \cite{10.1109/ICSE-C.2017.8}, layout mining \cite{10.1145/3126594.3126651} and accessibility analysis \cite{10.1145/3377811.3380392}, etc.
Yet the UI hierarchy generated by UIAutomator only contains limited information.
Specifically, whether a component can be included in the UI hierarchy is determined by the component's corresponding view internal attributes, e.g. alpha, invisible. When a component is transparent, invisible or beyond the region of the screen, it will be eliminated by UIAutomator.
Thus only a subset of the whole components in a UI page is obtained.

By comparison, Layout Inspector can be used for collecting the \textbf{view tree} including all views which are regardless visible or not, their images, and other attributes that UIAutomator can obtain \cite{Using_Views, Layout_Inspector}.
With Layout Inspector, we could collect a more complete view tree and the corresponding image of each view inside. 
Therefore, this paper utilizes Layout Inspector for better analyzing the apps and the follow-up inconsistency detection.

\section{Motivational Study}\label{UIAutormator_and_Layout_Inspector}

To better understand the scaling issues in real-world practice, we carry out a pilot study to examine the prevalence of these issues. 
The pilot study also explores what kinds of scaling issues exist, so as to facilitate the design of our approach for detecting UIs with  scaling issues.

\begin{figure*}[h]
  \centering
  \includegraphics[width=17.5cm]{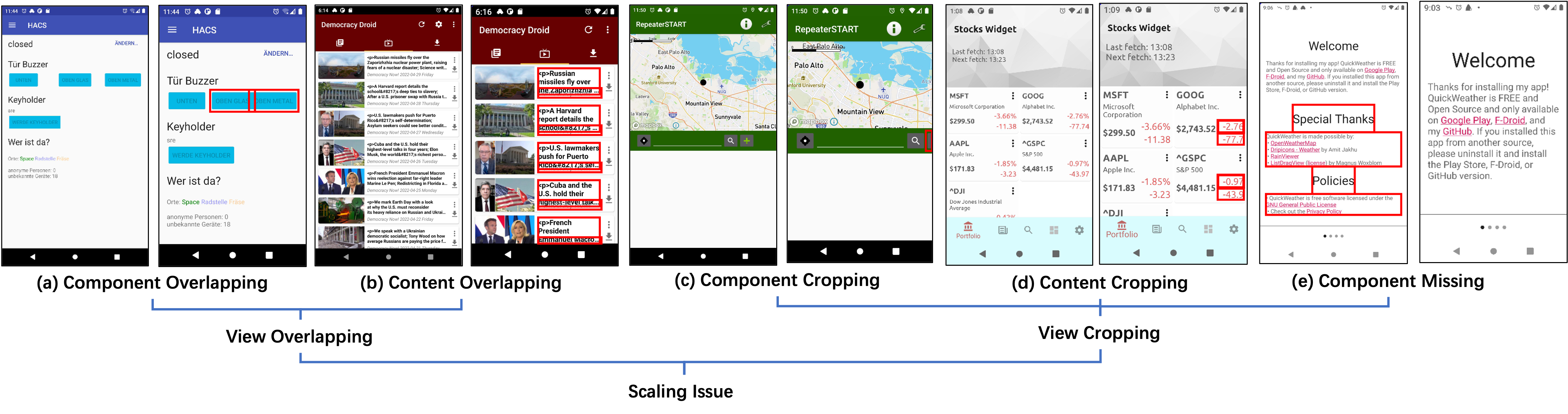}
   \caption{Examples of five categories of scaling issues. For each pair, the left image comes from the device with the default display scale, and the right one is from the larger display scale.
   }
  \label{fig:sri}
\end{figure*}

\subsection{Data Collection}
\label{subsec_motivational_data}

Our experimental dataset is collected from F-Droid \cite{f-droid}, a well-known open-source Android hosting website.
We employ the following criteria for app selection: more than 1K downloads in Google Play (popular), more than three years of development history (trustworthy), and more than 100 code commits (well maintained).
We then randomly select 200 apps for the pilot study. 
For each app, we record all the UI screenshots with the default display scale, and their counterpart screenshots with the largest display scale. 
In total, we collect 476 screenshots pairs.

\subsection{Categorizing Scaling Issues}
\label{subsec_motivational_issues}

Given those UI screenshots pairs, the first three authors individually check each of them manually, and determine whether the screenshot under the largest display scale has scaling issues. 
Only the screenshots with the consensus from all three human markers are regarded as ones with scaling issues. 
A total of 96 scaled screenshots are determined as having scaling issues, which accounts for 20.1\% (96/476) in all these scaled screenshots. 
These buggy screenshots come from 60 examined mobile apps, accounting for 30.0\% (60/200) of the examine mobile apps. 
These results indicate that the scaling issues account for a non-negligible portion of mobile apps, and should be paid careful attention to improving the software quality and accessibility. 

During the manual examination process, we notice that there are different types of scaling issues, a categorization of those issues would facilitate the understanding of them and the design of the approach. 
Following the Card Sorting \cite{spencer2009card} method, we classify those scaling issues into two categories with five sub-categories. 
We list the details as follows in Figure \ref{fig:sri}.
Note that, since there might be more than one scaling issue in a screenshot, the total percentage of the following categories is above 100\%.
\begin{itemize}[leftmargin=10pt,noitemsep,topsep=0pt]

\item \textbf{View Overlapping (24\%)}: two sibling views overlap with each other.

\begin{itemize}[leftmargin=5pt]
\item \textbf{Component overlapping (12\%)}: as shown in Figure \ref{fig:sri} (a), the buttons are overlapped together when using a larger display scale, and the content of button at the bottom cannot be seen completely. 

\item \textbf{Content overlapping (16\%)}: two pieces of text overlap each other. 
\end{itemize}

\item \textbf{View Cropping (80\%)}: the child view is blocked by its parent view, or by the view itself. 

\begin{itemize}[leftmargin=5pt]
\item \textbf{Component Cropping (20\%)}: the component is cropped or truncated under a larger display scale, e.g., the information is partly visible. 

\item \textbf{Content Cropping (60\%)}: part of the  information in a component is totally invisible, and the users might not be aware of the information loss. 

\item \textbf{Component Missing (23\%) }: some components are totally missing from the UI page (cannot be scrolled), thus the uses cannot reach certain functionalities. 

\end{itemize}
\end{itemize}

Note that, this categorization is only for demonstration, not for evaluation. 
Besides, there are other studies that categorize the UI related issues, e.g., \cite{DBLP:conf/kbse/LiuCWHHW20} categorize the UI display issues as component occlusion, text overlap, missing image, NULL value, and the blurred screen. 
The last three categories can hardly occur in our scenario, while the first two categories roughly correspond to the view cropping and view overlapping respectively. 
Yet we refine these categories to provide a more fine-grained perspective, also some sub-categories in this paper can not be detected by the approach of \cite{DBLP:conf/kbse/LiuCWHHW20}.

These issues can be triggered in terms of multiple views, e.g., the improper setting of the relative position of sibling views on a UI page, the size of the parent view is not large enough or unadjustable for hosting the child view, etc. 
They can also be caused in terms of a single view, e.g., truncated TextView because of the unadaptable setting. 
What's more, the involved types of views of each category of scaling issues are not limited to the Button or TextView as displayed in Figure \ref{fig:sri}. 
Every category of scaling issues can occur in a vast variety of views, and our website provides more examples for an intuitive understanding of these issues as well as the challenges in detecting them.

\subsection{Challenges in Scaling Issue Detection}
\label{subsec_motivational_challenge}

The above findings confirm the severity of scaling issues, and motivate us to design an approach for automatically detecting these scaling issues. 
We observe that some screenshots with scaling issues do not have clear visual characteristics, e.g., a screenshot with a component missing looks good if not compared with the screenshot of default scaling, thus cannot be automatically detected with the visual understanding techniques like OwlEyes \cite{DBLP:conf/kbse/LiuCWHHW20}.
On the other hand, these scaling issues can be detected by comparing the UI pages under default and larger display scale and checking the inconsistency. 
Yet there are the following three challenges in doing so. 

Since directly comparing two UI pages for scaling issue detection can easily bring much noise, the first challenge is building the mapping relations between two views under different display scales, which lays a solid foundation for the follow-up inconsistency detection. 
Previous approaches \cite{8115644} utilized the Android view id or XPath for the view mapping, but a large portion of view pairs can not be retrieved or retrieved not correctly, since the XPath of a view inside scrollable layout could change when scrolling this view, etc.
Second, the views within a page can interact with each other, since the size and position of a view within a page could be affected by other views, and when using a large display scale, the interaction effect makes the inconsistency detection more complex. 
Third, when using a larger display scale, different types of content (e.g., text, image) in a view might be scaled differently; thus the inconsistency detection should consider the characteristic of different types of views. 

\section{approach}

In this section, we present {\tool} to automatically detect the scaling issues.
The basic idea is to compare a VT under the default display scale and its counterpart VT under a larger display scale, and check the inconsistency between them. 

\begin{figure}[h]
  \centering
  \includegraphics[width=\linewidth]{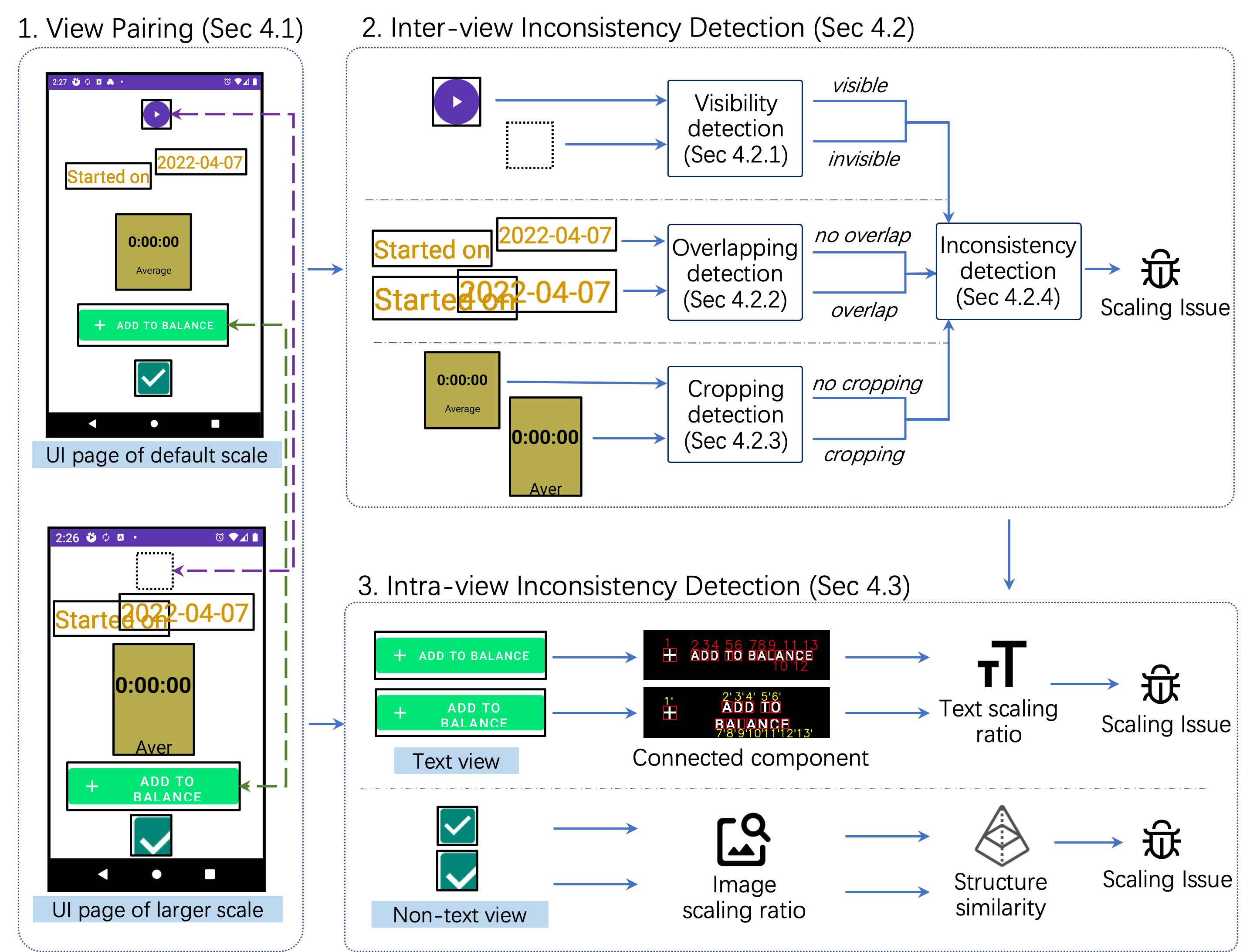}
  \caption{Overview of {\tool}. It inputs two UI pages (artificially created for demonstration) under default and larger display scale, and outputs whether each view has scaling issues. 
  }
  \label{fig:approach_diagram}
\end{figure}

Figure \ref{fig:approach_diagram} provides a high-level overview of this approach and shows its main phases.
The first phase is \textbf{view pairing}. 
For a VT under default display scale and its counterpart VT under larger display scale, we map each view with its  counterpart view for better conducting the follow-up inconsistency detection. 
The second and third phase checks the consistency from the perspective of the interaction between views and single view, respectively. 
In detail, the second phase is \textbf{inter-view inconsistency detection}, which detects whether there is any overlapping/cropping between two views in a VT, and checks whether the overlapping/cropping status keeps consistency for two VTs.
The third phase is \textbf{intra-view inconsistency detection}, which checks the inconsistency for each single view in a VT and its counterpart VT.

\subsection{View Pairing}

To mitigate the limitation of utilizing the android view id or XPath for retrieving the view pair as in previous approaches \cite{8115644}, we design a method to accurately find the view pairs by injecting unique id into views inside layout XML files, and finding the mapped views assisted by the textual information.

In detail, for an app, we decompile it, extract all of its layout XML files declaring UI elements, inject a unique mapping id for each XML element, recompile and get an ID injected app.
After this modification, we collect the VT corresponding to a specific UI page from this app with a UI execution script for executing apps to reach the page we want to test.
The view from the VT could contain our injected mapping id as its attribute.
Since the layout XML file is reusable, if a view is injected with a mapping id, we could obtain this view inside any VT when this layout XML file is used.
Based on this technique, we can not only obtain view pairs more precisely to facilitate this study, but also potentially guide the developers with the unique id in efficiently finding the corresponding buggy layout XML file for issue fixing.

For the case that a mapping id corresponds to a single view in a VT, we directly find the view with the same id in another VT (i.e., under a larger display scale), and determine them as a view pair. 
For the case that a mapping id corresponds to multiple views, e.g., views inside a scrollable list, we use the text information to assist the view pairing. 
In detail, 
for each repetitive view, based on its mapping id, we collect all of the text information of non-repetitive views inside this item to enhance the mapping id of the repetitive view to distinct this view from others. 
This is because inside those views, the text information of the off-springs
are usually unique and can be used for identifying the repetitive view. 
We first initialize the enhanced id of the repetitive view as a string of this repetitive mapping id. 
Then we traverse the VT in pre-order, and append the enhanced id with the mapping id and text information the current non-repetitive traversed view has. 
After traversal, we use the final enhanced id to distinguish it from other repetitive items.
In a VT, if two or more views share the same enhanced id, we assume those views sharing the same information, and just take one of them for mapping.
\subsection{Inter-view Inconsistency Detection}

This phase focuses on detecting the scaling issues induced by the interaction among views, in four steps.
In detail, we first check the view invisibility status (Section \ref{view_invisibility_detection}), which both helps reveal the inconsistencies and filters out the invisible views for better supporting the follow-up detection. 
Next, we check the overlapping between sibling views (Section \ref{Overlapping_detection}), and the cropping between the child view and its parent view (Section \ref{Parent_Cropping_Detection}).
We then detect whether there is inconsistency for the visibility, overlapping and cropping status between a VT and its counterpart VT (Section \ref{inter_view_inconsistency_Detection}).

\subsubsection{\textbf{View Visibility Detection}}\label{view_invisibility_detection}

\begin{figure}[t]
  \centering
  \includegraphics[width=\linewidth]{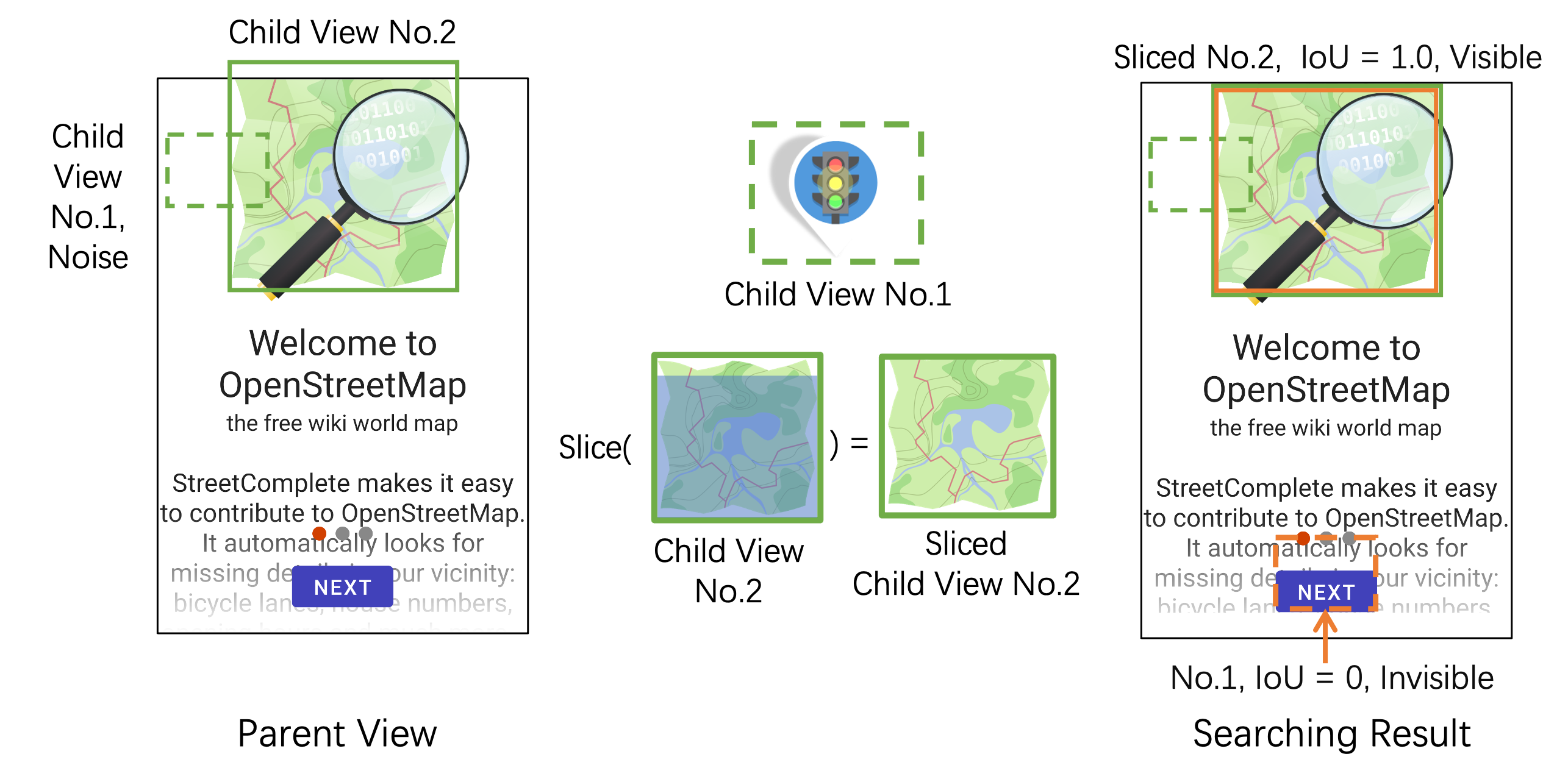}
  \caption{View visibility detection (Section \ref{view_invisibility_detection}).}
  \label{fig:template_matching_diagram}
\end{figure}

Before the inconsistency detection, we first need to check if a view is visible or not for end-users, and filter out those invisible noisy views to reduce the side effect of inconsistency detection.
To verify the visibility of the child view, we use a template matching algorithm \cite{DBLP:journals/cssp/YooH09} to check whether the image of the child view appears in the image of its parent view, as demonstrated in Figure \ref{fig:template_matching_diagram}.

We begin this phase from the root view of a VT, and initialize the root view as visible.
For a child view inside the parent view, we obtain the image of the child view and its parent view.
Since some child views might be sliced by the border of the UI page as the \textit{Child View No.2 in Figure \ref{fig:template_matching_diagram}}, we first obtain the sliced image $S$ of the child view for facilitating the accurate template matching.
In detail, we first obtain the bounding boxes of this child and its parent from the VT, based on this, we check if the child's bounding box is fully inside its parent by computing the union bounding boxes, and get the sliced child view accordingly.
We use the sliced child view $S$ as the query template, and the image of the parent view $T$ as the target to search on, then compute the searching result $R$.

Because the image of the parent view could have larger or the same size as the image of its child, the width and height of the searching result $R$ would be $(T.width - S.width +1)$ and $(T.height - S.height +1)$, respectively.
The template matching algorithm generates a corresponding score for each position inside $R$,
which indicates whether this position is the centroid of the searching result.
We treat the position with the highest score as the centroid of $R$ and obtain the corresponding result bounding box with the same shape of query image $S$. 
If the result bounding box has an Intersection over Union (IoU) \cite{DBLP:conf/cvpr/RezatofighiTGS019} as 1.0, the child view is visible; otherwise is invisible.
Only the visible views will be input into the following steps for further analysis.

\subsubsection{\textbf{Overlapping Detection Between Two Sibling Views}}\label{Overlapping_detection}

\begin{figure*}[!th]
\centering
\includegraphics[width=14.5cm]{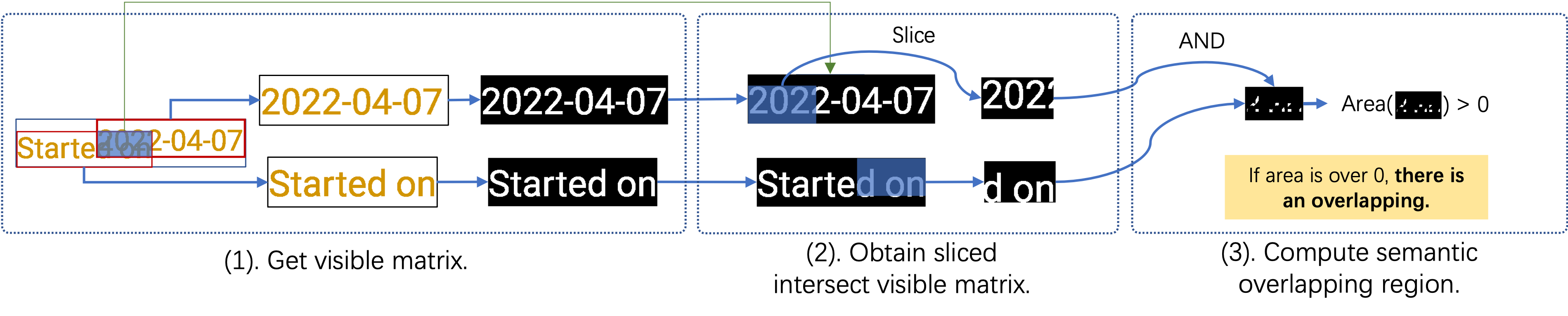}
\vspace{-0.1in}
\caption{Overlapping detection between two sibling views (Section \ref{Overlapping_detection}).}
\label{fig:overlapping}
\end{figure*}
For each visible view produced by the previous step,
there could be its siblings in a VT. 
Since its siblings could be drawn above it, some part of this view could not be visible, or it could overlap other siblings as shown in Figure \ref{fig:sri}.
Taken in this sense, this step detects the overlapping siblings of a view. 

The general practice of overlapping view detection is to find the overlap region of the bounding box for two involved views, yet this is not accurate since the real content of a view is not always demonstrated as a rectangle region.
In other words, the content inside the views might not be affected even if they have overlapping in the bounding box.
To tackle this, this paper designs a more fine-grained overlapping detection method, with three sub-steps as shown in Figure \ref{fig:overlapping}.

\textbf{\textit{Get visible matrix.}} To precisely determine whether two views have overlap in their content, we utilize the alpha channel to retrieve the region of content from the view's image, then conduct the follow-up overlapping detection based on the retrieved region. 
Specifically, the alpha channel of a view is a matrix that is of the same size as the view itself, and the value is between 0 and 255, where 0 indicates there is no content in the corresponding point. 
We, therefore, change the value which is larger than 0 to 1, and obtain a new matrix, which is called \textit{visible matrix $V$} to indicate the real content of a view.
In this matrix, 1 is for visible, and otherwise for 0. 
During inter-view analysis, this visible matrix will be kept for every view and might be updated following the overlapping detection results.

\textbf{\textit{Obtain sliced intersect visible matrix.}}
For two sibling views, we can obtain the Intersection over Union (IoU) \cite{DBLP:conf/cvpr/RezatofighiTGS019} of their bounding box, as the highlighted blue region in Figure \ref{fig:overlapping}.
According to this overlapping bounding box, we slice the overlapping region from the visible matrix of the two involved views.
By this, we obtain the sliced intersect visible matrix of the two views, which indicates the overlapped area of the real content in the two views.

\textbf{\textit{Compute semantic overlapping region.}}
We run the element-wise bit-wise AND operation between them and get a result matrix, i.e., if the value from the same position in these two matrices is both 1, the value of this position in the result matrix is 1; and 0 otherwise.
If the area of the result matrix is larger than 0, we assume there is an overlapping between the two sibling views.

Note that some overlapping views could be separated from each other by moving one of them.
For example, for DrawerLayout, even if it is drawn on another sibling, it won't fully block the view behind it because it can be collapsed.
Therefore, we do not consider this type of view for overlapping sibling detection.

We then update the visible matrix of the involved views to facilitate the follow-up detection. 
In detail, we subtract the result matrix from the sliced interaction visible matrix to get the subtraction matrix.
Then we replace the intersection region of the visible matrix of the involved views with the subtraction.

\subsubsection{\textbf{Cropping Detection between Parent View and Child View}}\label{Parent_Cropping_Detection}

After we analyze the overlapping between siblings, we move on to the next phase by investigating whether a view could be cropped by its parent, with three similar sub-steps as overlapping detection.

The general mechanism of cropping detection is similar to the overlapping sibling detection in the previous section. 
For the first two sub-steps, i.e., \textit{get visible matrix} and \textit{obtain sliced intersect visible matrix}, it utilizes the same processing method. 
For the third sub-step  \textit{compute semantic cropping region}, it obtains the area of the result matrix, and if the intersected area is smaller than the child view, we assume the child view is cropped by its parent view
After that, we update the visible matrix of the involved views to facilitate the follow-up detection in a similar way.

We also note that certain cropping views could be made visible through scrolling its parent view. 
We detect whether a view is scrollable with its class name, i.e., ScrollView, RecyclerView, and do not consider these views for cropping detection.

\subsubsection{\textbf{Inconsistency Detection in Terms of Two Views}}
\label{inter_view_inconsistency_Detection}

Based on the view visibility detection (in Section \ref{view_invisibility_detection}), overlapping and cropping detection between two views (in Section \ref{Overlapping_detection} and \ref{Parent_Cropping_Detection}), we check whether the visibility, and the overlapping/cropping status keep consistent among two VTs, and determine whether there is a scaling issue. 
For example, if a view overlaps with another view inside a VT, yet these two views don't overlap in its counterpart VT, there is a scaling issue for the view pair.

\subsection{Intra-view Inconsistency Detection}

After determining the consistency for the overlapping or cropping status between views, we then shift our focus on the visible leaf view, and check the inconsistency of a leaf view between a VT and its counterpart VT. 

In this phase, we separate views into two categories, i.e., text view and non-text view, and design specific methods accordingly. 
If a view has text attributes, we treat it as a text view; otherwise, we treat it as a non-text view. 
Note that, there can be icons in a text view, and there can also be text information in a non-text view, which complicates the inconsistency detection. 

\begin{figure*}[h]
\centering
\includegraphics[width=16cm]{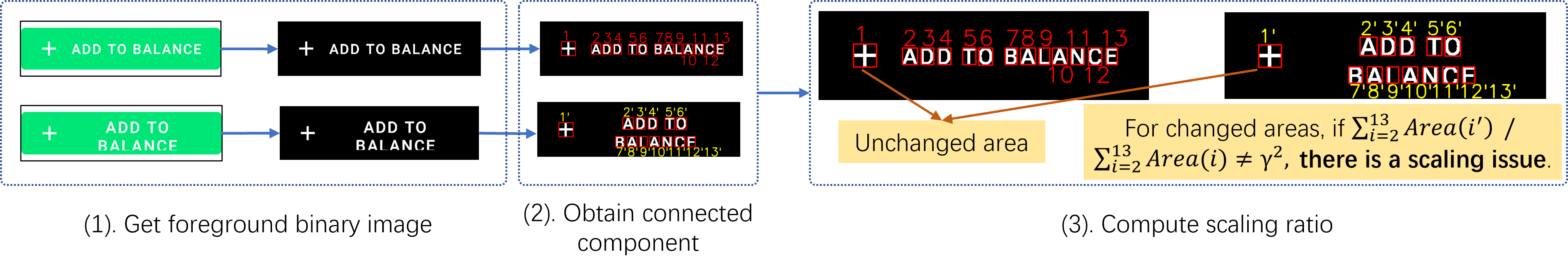}
\vspace{-0.15in}
\caption{Inconsistency detection in terms of single text view (Section \ref{Cropping_Detection_for_Single_Text_View}). 
Inputs are two views under default and a larger display scale. 
Each white part is a connected component, marked with its index number and bounding box. 
}
\label{fig:connected_component}
\end{figure*}

\subsubsection{\textbf{Inconsistency Detection in Terms of Single Text View}} \label{Cropping_Detection_for_Single_Text_View}

We compare two text views in a view pair by analyzing if the texts are scaled properly in terms of the area of textual information. 
The common practice of comparing two text-rich images is utilizing the Optical Character Recognition (OCR) technique \cite{mori1992historical}, yet it needs loads of training data and is far from being accurate when faced with complicated or infrequent combination of characters or words. 
This paper employs a finer-grained method, i.e., recognizing the connected components from the image, which captures the semantic content of the image. 
Connected components detection is a widely used algorithm in image processing for region extraction  \cite{DBLP:journals/ejivp/SujathaS15,DBLP:journals/tcs/FiorioG96,DBLP:conf/miip/WuOS05}.

\textbf{\textit{Get foreground binary image.}}
We first need to retrieve the foreground of the text view, i.e., removing the influence of background to the following-up detection.
This is done through thresholding the image of the view. 
Image binarization replaces pixels in a grayscale image where the value is larger than a threshold with a black pixel (0), and replaces pixels where the value is smaller than a threshold with a white pixel (255) \cite{shapiro2001computer}. 
We set the threshold with Otsu's method, which performs automatic image thresholding with the content of the grayscale image, separating pixels into foreground and background \cite{liu2009otsu}.
After this, we obtain a binary image of the foreground containing texts (and icons if any).

\textbf{\textit{Obtain connected components.}}
Figure \ref{fig:connected_component} demonstrates how we compare the foreground binary images of a pair of text views by retrieving the connected components.
The connected component is generated by segmenting a binary image with the values inside \cite{DBLP:conf/miip/WuOS05}. 
For each value of 255 (white part) inside a binary image, if it has a neighbor which is also 255, then they belong to the same connected component, and we can obtain the bounding box and area of each connected component. 

\textbf{\textit{Compute scaling ratio.}}
With the region of each connected component, we can sum up the area for all the connected components in an image. 
If a text view is scaled to $\gamma$ times compared with its counterpart view, the ratio between the area of these two views would be $\gamma^2$ (i.e., the width and the height of each pixel is scaled to $\gamma$ times).
Taken in this sense, if the ratio does not satisfy this, a scaling issue exists. 
However, if the text inside is ellipsized, which is determined by the ``getEllipsize'' attribute, we would ignore it.

Note that, when scaling a view, the icon might not be affected, e.g., icon \textit{1 and 1'} as shown in Figure \ref{fig:connected_component}, when only changing the font size.
To reduce their impact on the result, when summing the area of connected components, we first eliminate these connected components which remain unchanged.

Since different views can demonstrate slight different scaling even under the same setting, to precisely determine the $\gamma$, we utilize the ``textSize'' attribute of a view in the VT, which shows the text size of a view in the current setting.
We obtain the ``textSize'' $size$ of a view and the ``textSize'' $size'$ of its counterpart view, then  $\gamma$ is computed as $size' / size$ for the view pair.

\subsubsection{\textbf{Inconsistency Detection in Terms of Single Non-Text View}}

For the non-text view, we combine the number of connected components and structure similarity to detect the inconsistency, because the non-text view is more complex and the fine-grained technique for handling text views in the previous section might fail to work in this situation.

In detail, we first obtain the foreground binary image of the view following the method described in previous section.
Since different elements in a non-text view can be scaled quite differently, we do not use the scalable area as that in the previous section for the inconsistency detection; rather we count the number of connected components, which is more  coarse-grained.
If the count is different between a VT and its counterpart VT, we think it has the risk of having a scaling issue, and use the structure similarity for further checking.

We employ the Structure Similarity Index Measure (SSIM) \cite{DBLP:journals/tip/WangBSS04} to further determine whether the looking of two views look similar. 
In detail, for each RGB channel of two views, we compute the SSIM value, and then obtain the average value of the three SSIM values and treat it as the similarity score. 
If the RGB channels of view are fully 0, we compute the SSIM value based on the alpha channel for the similarity score. 
If the similarity score is under a pre-defined threshold (set as 0.9 empirically), we report a scaling issue. 

\subsection{Implementation}

For the VT collection, we employ the ddmlib \cite{ddmlib}, which is a tool for retrieving VT and corresponding images of an app from Android runtime information. 
For injecting a unique id for the views, we first decompile an apk with apktool \cite{apktool} and get its decompiled result folder containing all the apk's internal files. 
We then obtain the layout XML files from the ``res/layout'' folder. 
For each file, we would inject the namespace of ``xmlns:android'' into the root element if there is none. Then we inject the mapping attribute ``android:mappingID'' into all nodes inside, and set a unique value for each of them. 

We also add ``<attr name=``android:mappingID''/>'' into the root node of ``res/values/attrs.xml''.
We would record all the mapping id we used and where it is used, for identifying where the buggy view comes from. 
After all the layout files are modified, we compile all the resources and get an ID injected apk.
For collecting VTs, we turn on the ``Enable view attribute inspection'' option in the developer settings of the emulator, making sure that VTs contain the mapping id attribute we injected in.
For each app we test, we generate a UI execution script for manipulating app on different devices and testing the same page.

We implement {\tool} with Python \cite{python}, and we use Numpy \cite{numpy} for manipulating matrices and OpenCV\cite{opencv} for reading images, retrieving binary images, computing connected components in binary images, etc. 
For the template matching algorithm in Section \ref{view_invisibility_detection}, we employ the ``TM CCOEFF NORMED'' matching strategy \cite{opencv_template_matching} in OpenCV, which contains two rounds of normalization and is robust to noise.

\section{Experimental Design}

\subsection{Research Questions}

\begin{itemize}[leftmargin=10pt]
  \item \textbf{RQ1: (Issue Detection Performance)} What is the performance of {\tool} for detecting \textit{UI pages} containing scaling issues?

  \item \textbf{RQ2: (Issue Localization Performance)} How effective is our {\tool} in localizing \textit{views} with scaling issues?

  \item \textbf{RQ3: (Usefulness Evaluation)} How effective is our proposed {\tool} in detecting scaling issues in practical apps?
\end{itemize}

We utilize RQ1 and RQ2 to investigate the performance of our proposed {\tool} in different granularity, i.e., buggy UI page and buggy view in the page, which also facilitate the comparison with different baselines. We also investigate the time cost to demonstrate its effectiveness further. 
RQ3 uses our proposed {\tool} to detect the scaling issues in real-world apps, and issue the detected bugs to the development team for demonstrating its effectiveness in real-world practice.

\subsection{Experimental Setup and Dataset}\label{ExperimentalSetupandDataset}

We use Appium \cite{Appium} to replay actions from UI execution script upon an app on a mobile device under 3 different scale settings: \textbf{d}efault display size and \textbf{d}efault font size (short for DD-size), \textbf{l}argest display size and \textbf{d}efault font size (LD-size),
\textbf{l}argest display size and \textbf{l}argest font size (LL-size).
We use Pixel 2 emulator as our target device, whose screen resolution is 1080$\times$1980.
After the replay is done, we gather all the VTs and the image of each view under each setting. 
Meanwhile, we collect the corresponding view hierarchy files and screenshots, which will be used for the baseline approaches. 
We run experiments  on an Intel Core i5-11400 machine, with 16GB of memory and installed with Ubuntu Linux 16.04 LTS.

To build the ground truth for evaluation, we manually label each view inside VT of LD-size (and LL-size) as buggy or clean compared with DD-size by checking whether there are any inconsistencies.
During the labeling process, three annotators individually conduct the manual checking, and the disagreement is sent over to one author for the final decision. 
Then a face-to-face meeting is organized for discussing the difference among the results, until a consensus is reached. 
This part of dataset is \textbf{view dataset}, and the following datasets are built with the same protocol of manual labeling. 
Besides, we label each UI component inside view hierarchy files of LD-size (and LL-size) as buggy comparing with NN-size if there are any inconsistencies. 
This part of the dataset is \textbf{hierarchy dataset}, and will be used for the baseline DiffDroid.
Also, we label each screenshot of LD-size (LL-size) as buggy comparing with NN-size in a similar way. This part of dataset is \textbf{screenshot dataset}, and will be used for the baseline OwlEyes and DiffDroid.

We reuse the 60 apps from the motivational study in Section \ref{subsec_motivational_data} for the evaluation of {\tool}.
Note that, this would not influence the experimental results since our {\tool} is not the learning-based approach which needs the separated training dataset and test dataset.
Based on them, we collect a \textbf{view dataset} with 213 clean VTs and 96 buggy VTs, with 23,663 views and 147 of them are buggy, a \textbf{hierarchy dataset} with 213 clean hierarchy XML files and 64 buggy files involving 133 buggy UI components, and a \textbf{screenshot dataset} with 213 clean screenshots and 96 buggy ones. 

\subsection{Evaluation Metric}\label{metric}

We utilize the commonly-used precision, recall, and F1-score to measure the performance. For all the metrics, a higher value represents better performance.
For the performance of detecting buggy UI pages (RQ1), the precision is the proportion of buggy UI pages that are correctly predicted among all UI pages.

Recall is the proportion of UI pages that are correctly predicted among all UI pages that really have scalling issues.

F1-score (F-measure or F1) is the harmonic mean of precision and recall, which combines both of the two metrics above.

For the performance of detecting buggy views (RQ2), the performance metrics are computed similarly. 
In addition, we also present the precision, recall, and F1-score for the clean category, i.e., clean UI page and clean view, which are computed in a similar way.

\subsection{Baselines}

\textbf{OwlEyes} \cite{DBLP:conf/kbse/LiuCWHHW20}, the state-of-the-art approach for detecting UI display issues, e.g., text overlap, component occlusion in a UI screenshot. 
It does not require two counterpart UI pages for comparing the inconsistency; by comparison, it builds on the Convolutional Neural Network (CNN) to directly predict whether a screenshot  has  UI display issues. 
For our experiment, we employ the trained model provided on its website, and input the screenshot under a larger scale for determining whether the scaling issue exists, i.e., UI display issue exists. 
Note that, OwlEye can also provide the issue localization results with 
Gradient weighted Class Activation Mapping technique in the screenshot. 
Yet it can only highlight a general region, rather than the detailed views as provided by our approaches, thus we only utilize this baseline in buggy UI page detection (RQ1), while do not use it for buggy view localization (RQ2).

\textbf{DiffDroid} \cite{8115644}, the state-of-the-art approach for detecting the cross-platform UI inconsistency, e.g., the display of the same page across different mobile devices. 
It extracts several features which indicate the visual similarity/difference of a pair of components from two UI pages, e.g., structural similarity, distance of color histogram, and builds a decision tree for predicting whether two UI pages have any inconsistent components. 
For our experiment, we reuse the code provided on its website, and collect the training and testing data with DiffDroid following its original paper.
In detail, we feed our hierarchy dataset and screenshot dataset into DiffDroid, get the list of component pairs and a set of features.
We also obtain whether a component  pair involving an inconsistency based on the labeled views in Section \ref{ExperimentalSetupandDataset}. 
Based on this labeled dataset, we utilize five-fold cross validation to obtain its performance.
This baseline can be applied for both buggy UI page detection (RQ1), and buggy view localization (RQ2).

\section{Results and Analysis}\label{Results_and_Analysis}

\subsection{Issue Detection Performance (RQ1)}\label{RQ1}

Table \ref{tab:RQ1_page_detection} presents the performance of {\tool} in detecting the UI pages with scaling issues.  
For bug type in Table \ref{tab:RQ1_page_detection} (i.e., pages having scaling issues), with {\tool}, the precision is 0.97, indicating 97\% of the page which are predicted as having scaling issues are correct. 
The recall is also 0.97, indicating 97\% of the pages with scaling issues can be found with our proposed approach.
For clean type (i.e., UI pages without scaling issues), the precision is 0.92 and the recall is 0.92, implying most of the clean page would not be misclassified. 
This indicates the near-perfect performance and the potential usefulness of scaling issue detection by our proposed approach.

\begin{table}[!ht]
\renewcommand\arraystretch{0.9} 
\caption{Performance of detecting buggy and clean pages.}
\vspace{-0.1in}
\label{tab:RQ1_page_detection}
\begin{tabular}{c|c|c|c|c}
\hline
\textbf{Method}&\textbf{Type}&\textbf{Precision}&\textbf{Recall}&\textbf{F1-Score} \\\hline
\multirow{2}{*}{dVermin} & Bug&\textbf{0.97}&\textbf{0.97}&\textbf{0.97}\\
\cline{2-5}&Clean&\textbf{0.92}&\textbf{0.92}&\textbf{0.92}\\\hline
\multirow{2}{*}{OwlEyes} & Bug & 0.39 & 0.50 & 0.44\\
\cline{2-5}&Clean& 0.78 & 0.69 & 0.73\\ \hline
\multirow{2}{*}{DiffDroid} & Bug & 0.30 & 0.11 & 0.16\\
\cline{2-5}&Clean& 0.71 & 0.90 & 0.80\\ \hline
\end{tabular}
\end{table}

Table \ref{tab:RQ1_page_detection} also demonstrates the baseline approaches OwlEyes and DiffDroid in detecting UI pages with scaling issues. 
We can see that for both bug type and clean type, our proposed approach outperforms the OwlEyes by a large margin, i.e., 148\% (0.97 vs. 0.39) higher in precision, 94\% (0.97 vs. 0.50) higher in recall for bug type. 
For DiffDroid, our proposed approach also outperforms by a large margin, i.e., 223\% (0.97 vs. 0.30) higher in precision, 781\% (0.97 vs. 0.11) higher in recall for bug type.
OwlEyes is designed to spot the UI display issues by the visual understanding, and is good at detecting the display issues with obvious visual characteristics, e.g., a component is occluded by another component. 
Yet in our scenario of UI page under larger scale, there might be a missing component or textual information as shown in Figure \ref{fig:sri}, which could not be directly detected by visual information. 
Furthermore, there might be some cases that look like having display issues and are detected as having issues by OwlEyes, yet the involved views can be scrolled (i.e., not an issue). 
We will present the reason for the bad performance of DiffDroid in the next section.

We then present the time cost of each approach for buggy page detection, to further demonstrate the effectiveness of our approach; and the time cost for buggy view detection shows the same trend.
For processing one pair of UI pages, it takes {\tool} a median of 70 seconds. 
Among them, about 97\% time is consumed for analyzing the inter-view analysis (Section \ref{inter_view_inconsistency_Detection}) which contains complex matrix computation. 
Nevertheless, for a typical app with five UI pages, the estimated time for the scaling issue detection is about 6 minutes (5 $\times$70 seconds), which is acceptable. 
For the baseline OwlEyes, the median time used for detecting a screenshot is 0.09s, yet training the model requires about 8 hours (mentioned in the original paper  \cite{DBLP:conf/kbse/LiuCWHHW20}).
For detecting a pair of UI pages with the baseline DiffDroid, the median time is 42 seconds.
Our proposed approach consumes comparable time (70 vs. 40 seconds), yet achieves far higher performance.

\subsection{Issue Localization Performance (RQ2)}\label{RQ2}

Table \ref{tab:RQ2_page_detection} presents the performance of {\tool} in localizing views with scaling issues.  
We can see that the precision is 0.84, and the recall is 0.91, indicating most of the views with scaling issues can be found with our proposed approach.
This indicates our approach is good at localizing the buggy view from the buggy UI page, which can further facilitate the detailed root cause analysis and bug fixing. 

\begin{table}[h]
\renewcommand\arraystretch{0.9} 
\caption{Performance of localizing buggy and clean views.}
\vspace{-0.1in}
\label{tab:RQ2_page_detection}
\begin{tabular}{c|c|c|c|c}
\hline
\textbf{Method}&\textbf{Type}&\textbf{Precision}&\textbf{Recall}&\textbf{F1-Score} \\\hline
\multirow{2}{*}{dVermin} & Bug&\textbf{0.84}&\textbf{0.91}&\textbf{0.88}\\
\cline{2-5}&Clean&\textbf{0.99}&\textbf{0.99}&\textbf{0.99}\\\hline
\multirow{2}{*}{DiffDroid} & Bug & 0.11 & 0.11 & 0.11\\
\cline{2-5}&Clean& 0.98 & 0.98 & 0.98\\ \hline
\end{tabular}
\end{table}

Table \ref{tab:RQ2_page_detection} also shows the performance of the baseline approach DiffDroid.
Our approach outperforms DiffDroid by a large margin, i.e., 664\%(0.84 vs. 0.11) higher in precision, 727\%(0.91 vs. 0.11) higher in recall for localizing buggy views.
According to our observation, we find the following  reasons for our promising results.
First, we design an accurate view pairing method with  injected mapping id, yet DiffDroid employ XPath to find the pair which is less accurate. 
Second, our approach can obtain the image of each view with Layout Inspector, while DiffDroid generates the image by slicing the screenshot by which the image could involve other views, thus induce noise. 
Third, since its original task is for the cross-platform inconsistency detection, and for each app, they have large amount of data collected from different platforms for model training, i.e., they experiment with 130 different platforms (e.g., mobile devices).
In detail, for each app, they use a portion of the platforms for collecting the training data, and use the remaining platforms for evaluating the performance. 
In other words, when conducting prediction, the model has learned the knowledge in terms of the specific app, and thus can achieve relatively high performance as the paper reported. 
By comparison, in our cross-scale scenario, since there are only limited scale settings, the performance is far from satisfactory compared with the cross-platform scenario.

\begin{figure}[h]
\centering
\includegraphics[width=1\linewidth]{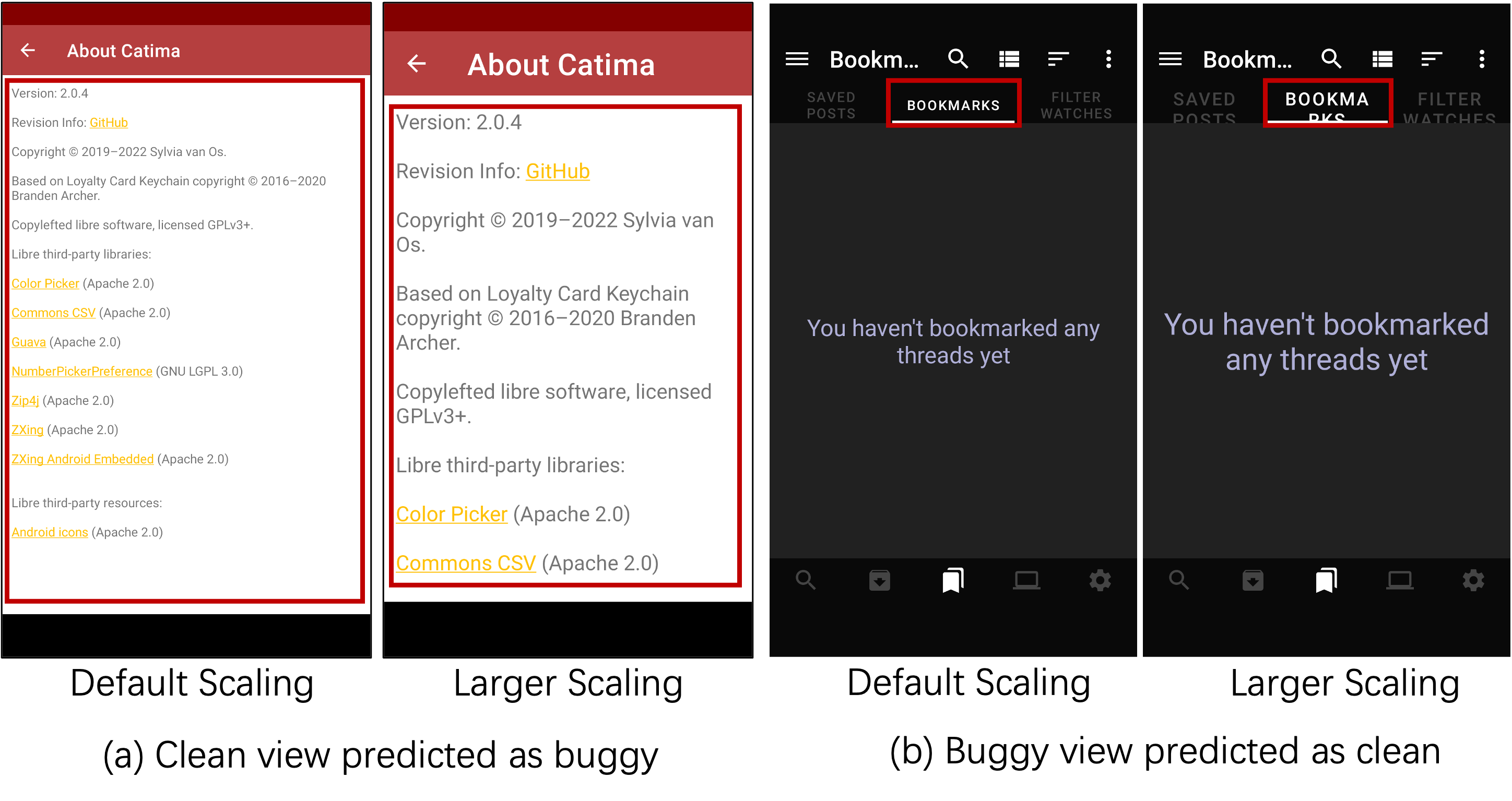}
\vspace{-0.3in}
\caption{Bad cases of {\tool}.}
\label{fig:bad_2}
\end{figure}
\vspace{-0.1in}

We further analyze the failure case of {\tool} in localizing views with scaling issues. 
In Figure \ref{fig:bad_2} (a), the view under the larger display scale does not have any scaling issue, yet wrongly be predicted as with scaling issues (i.e., cropped), because {\tool} finds the  inconsistency between the two views. 
However, the text view is scrollable; thus there is no issue. 
Our approach checks whether a view is scrollable according to the type of the view (e.g., ScrollView, RecyclerView). 
Since some scrollable views may be set dynamically when running the code, it is beyond the scope of this study.
For Figure \ref{fig:bad_2} (b), the view with a scaling issue is wrongly predicted as a clean view. 
The view has ``getEllipsize'' attribute, and {\tool} determines it can be with ellipsis (i.e., not treated as scaling issue), but its appearance is cropped. 
Future work would employ more strategies to handle these corner cases to further improve the performance.

\subsection{Usefulness Evaluation (RQ3)}\label{RQ4}

To further assess the usefulness of our {\tool}, we use the criteria in Section \ref{subsec_motivational_data} to sample the apps from F-Droid, and randomly sample 80  Android applications.
We utilize {\tool} to detect scaling issues for these apps. 
Once a scaling issue is detected, we create a bug report by describing the issue attached with a buggy UI screenshot to better illustrate the issue. 
Finally, we report them to the app development team through issue reports.

Table \ref{tab:issues} presents the details of the submitted scaling issues. 
In total, 21 scaling issues are detected, and 20 (95\%) of  them have been confirmed/fixed by app developers, while the remaining one is pending.
These fixed or confirmed issue reports further demonstrate the effectiveness and usefulness of our proposed approach in detecting scaling issues. 
The developers also express thanks in the issue's comments. 
For example, one developer from WHWDataset (No.4) left a comment as ``\textit{Thanks a lot! Accessibility is always one of those tricky things that's hard to cover unless we spend a lot of time testing}''. 
And another developer from EVmap (No.5) left a comment ``\textit{This is definitely a bug, I didn't test these intro screens with so large scaling and font size}''.

\begin{table}[h]
\caption{GitHub submitted issue reports.}
\renewcommand\arraystretch{0.9} 
\vspace{-0.1in}
\label{tab:issues}
\centering
\resizebox{\columnwidth}{!}{
\begin{tabular}{c|c|c|c|c|c|c}
\hline
\textbf{No.} & \textbf{Issue Link ID} & \textbf{APP Name} & \textbf{Category} & \textbf{Version} & \textbf{Download}&\textbf{Status} \\
\hline
1 & 304671 & GPSLogger & Navig & 112 & 1M+&fixed \\
2 & 810349 & AlarmClock & Time & 3.9.1 & 1M+&fixed \\
3 & 624992 & PDFConverter & Tool & 8.8.1 & 500K+&fixed \\
4 & 518169 & WHWDataset & Tool & 7.1.5 & 500K+&fixed \\
5 & 572599 & EVMap & Health & 0.8.3 & 500K+&fixed \\
6 & 760709 & EinkBro & Internet & 8.12.1 & 50k+&fixed \\
7 & 208094 & Easer & Internet & 0.8 & 50K+&fixed \\
8 & 564646 & BabyDots & Tool & 1.5 & 50K+&fixed \\
9 & 652416 & Antimine & Tool & 12.4.2 & 50K+&fixed \\
10 & 494843 & BCWallet & Finance & 8.16 & 50K+&confirmed \\
11 & 428996 & TickMate & Health & 1.4.12 & 10K+&fixed \\
12 & 211135 & PicardCode & Media & 1.5 & 10K+&fixed \\
13 & 797738 & PleesTracker & Tool & 7.1.5 & 10K+&fixed \\
14 & 876101 & Silence & Navig & 1.6.2 & 10K+&fixed \\
15 & 822374 & WeeklyBudget & Finance & 1.4 & 10K+&fixed \\
16 & 329138 & OpenWeb & Internet & 0.3 & 10K+&fixed \\
17 & 577533 & DemoDroid & Social & 4.2 & 10K+&fixed \\
18 & 546964 & Badreads & Reading & 0.1.6 & 10K+&fixed \\
19 & 701634 & APKEditor & Tool & 0.17 & 10K+&fixed\\
20 & 544654 & UMLEditor & Tool & 1.0 & 10K+&fixed\\
21 & 502273 & OSMTracker & Tool & 1.0.1 & 10K+&pending\\
\hline
\end{tabular}
}
\vspace{-0.2in}
\end{table}
\section{Threats to Validity}

The external threats concern the generality of the proposed approach. 
We evaluate {\tool} on 60 mobile apps from various categories involving finance, navigation, media, etc. 
These apps are all commonly-used and popular ones, which relatively reduces this threat.
In addition, we demonstrate the usefulness of {\tool} in detecting real-world scaling issues with affirmative results. 

Regarding internal threats, we reuse the trained model of the baseline OwlEyes; reuse the source code of the baseline DiffDroid, and strictly follow the description in the original paper for model training. 
This helps ensure the accuracy of the experiments.

The construct validity of this study mainly questions of the experimental setup of our approach.
For each UI page, we collect the setting under three display scales, i.e., default display and font size, largest display size and default font size, the largest display and font size, considering the fact that  combining the display and font size can complicate the UI display.  
This can help diversify our experimental dataset and better evaluate the robustness of our approach. 

\section{related work}

Graphical User Interface (GUI) is critical to the app accessibility, and related studies included GUI code generation \cite{nguyen2015reverse, chen2018ui, moran2018machine, chen2019storydroid}, GUI changes detection and summarization \cite{moran2018automated, moran2018detecting}, automatic GUI search \cite{10.1145/3359282}, etc.

Mobile accessibility issues like accessibility property missing or small touchable size of components could impede the usability of apps \cite{DBLP:conf/icse/Alshayban0M20, 10.1145/3491102.3502143}.
Many Android accessibility tools
contributed to solving accessibility issues by detecting and identifying known issues based on violations of guidelines, e.g., Accessibility Scanner \cite{accessibility_testing}, UIAutomator \cite{accessibility_testing}, Lint \cite{accessibility_testing}.
Espresso \cite{accessibility_testing} and Robolectric \cite{accessibility_testing} are frameworks for running testing assertions specified by developers for interacting with applications to test the accessibility.
MATE \cite{DBLP:conf/icst/ElerRGF18} analyzed visual accessibility issues by exploring apps automatically and checking accessibility properties as content descriptions and the contrast ratio of UI components. 
Latte \cite{10.1145/3411764.3445455} could reuse test cases, detect and generate functional accessibility warnings with SwitchAccess, e.g., unfocusable but touchable buttons or untouchable buttons with focusable attributes.
This study targets at another type of accessibility issue to help improve the usability of apps.

To ensure the quality of Android GUI, there are approaches for detecting UI display-related issues, e.g., display issues \cite{DBLP:conf/kbse/LiuCWHHW20,su2021owleyes,10.1145/3276526,10.1145/3238147.3238180}, cross-platform inconsistency \cite{8115644}, UI rendering problems \cite{gao2017every,li2019Characterizing}, UI performance differences \cite{DBLP:conf/icse/KiPDKZ19}, setting-related defects \cite{DBLP:conf/issta/SunSLDPXS21}, etc, and tools \cite{liu2022Guided,liu2022navidroid} for assisting testers to find UI display-related issues.
None of these techniques can be applied in our scaling issue detection.

There are also techniques for detecting the inconsistency between a pair of web pages in different web browsers, e.g., \cite{mahajan2015websee,mahajan2016using,DBLP:conf/icse/MahajanAMH18,5609723,2012CrossCheck}.
Typically, they utilized computer vision and image processing techniques to compare a web page with an oracle image (under default setting) to find the  visual differences. 
Different from web apps, Android apps have a different rendering mechanism, and it is difficult to analyze the fine-grained display status of Android apps than the web apps written in JavaScript, thus we propose this new approach. 

\section{conclusion}

Although users always scale up the font or display size of GUI for better readability, little is known about the scaling issues induced by the scale change. 
We first conduct a pilot study to examine the scaling issues in real-world apps, and then propose an automated approach {\tool} for detecting these issues in mobile apps. 
Experimental evaluations show its promising performance in precision and recall, and its usefulness in detecting real-world scaling issues with confirmed/fixed issue reports by developers. 

Future work will extend {\tool} to handle inconsistencies dynamically introduced when the app is running, and more evaluation on more apps and more scale settings.
We will also explore the root causes of these scaling issues and design corresponding automatic repairing techniques to help developers improve the accessibility of their mobile apps.

\begin{acks}
This work is supported by the National Key Research and Development Program of China under grant No.2018YFB1403400, National Natural Science Foundation
of China under Grant No. 62072442, No. 62002348, and Youth Innovation Promotion Association Chinese Academy of Sciences.
\end{acks}

\balance
\bibliographystyle{ACM-Reference-Format}
\bibliography{sample-base}

\end{document}